# INTEGRATING GOALS AFTER PRIORITIZATION AND EVALUATION – A GOAL-ORIENTED REQUIREMENTS ENGINEERING METHOD

Vinay S[1], Shridhar Aithal[2] and Sudhakara Adiga[3]

[1]Department of Computer Science & Engineering, PESITM, Shimoga, India
[2]Advanced Tech. Consultant, Canada
[3]Department of Mathematics, MIT, Manipal University, India

## ABSTRACT

*Decision support system in Requirements engineering plays an important role in software development life cycle. The relationship between functional and non-functional requirements often plays a crucial role in resolving conflicts or arriving at decisions in requirements engineering phase. Goal-Oriented Requirements Engineering (GORE) methods make a good attempt of addressing these aspects which are helpful in decision support. We propose a GORE method - Integrating goals after prioritization and evaluation (IGAPE). The method is semi-formal in nature thereby ensuring active stakeholder participation. In this paper we elaborate the various steps of IGAPE method. The output of IGAPE is then given as input to a decision support system which makes use of Analytic Hierarchy Process (AHP) and Technique for Order of Preference by Similarity to Ideal Solution (TOPSIS). Integration of IGAPE with AHP and TOPSIS will clearly provide a rationale for various decisions which are arrived at during the requirements engineering phase. The method is illustrated for an e-commerce application and is validated by expert analysis approach.*

## 1. INTRODUCTION

Requirements engineering phase of software development life cycle involves identifying functional and non-functional requirements. While functional requirements are important, eliciting and capturing the non-functional requirements (NFR) during the requirements engineering phase becomes even more important [1]. Traditional systems analysis approaches [2],[3] and [4]) during requirements analysis treat requirements as consisting only of processes and data and do not capture the rationale for the software systems. This makes it difficult to understand requirements with respect to high-level goals in the problem domain. Traditional modeling and analysis techniques do not allow alternative system configurations to be represented and evaluated [5]. Non-functional requirements Goal-oriented requirements engineering (GORE) approaches make a good attempt to address the essential quality characteristics which are commonly known as non-functional requirements ([6] and [7]). Benefits associated with elicitation, refinement and analysis of goals are addressed in [7].

Gunther Ruhe [8] highlights the importance of Decision Support system (DSS) in Software Engineering. Decisions are the driving engines for all stages of software development and evolution. Decisions can be related to methods, tools, and techniques. Decisions are aimed at answering the questions 'How'? 'How good'? 'When'? 'Why'? and 'Where'?. The objective is to have a sound methodology which provides rationale for the decision arrived at. The importance of decision making techniques is also addressed in ([9], [10] and [11]). The importance of stakeholders in decision making is done by Jenny P. and Jacob C [12]. Architecture decision making is closely linked to requirements engineering and the aspects related to this are addressed in Ana Ivanović and Pierre America [13] and Vidya Lakshminarayanan et al [14].

Decision-Making in Software Engineering is extremely challenging because of a dynamically changing environment, conflicting stakeholder objectives, constraints, coupled with a high degree of uncertainty and vagueness of the available information. While it is important to gather the NFR, it is also equally important how each of NFR is affecting the overall system goals.

We propose a GORE method - Integrating goals after prioritization and evaluation (IGAPE) in this paper. The output of IGAPE is then used as input to decision support system which makes use Analytic Hierarchy Process (AHP) and Technique for Order of Preference by Similarity to Ideal Solution (TOPSIS) methods.

Analytic Hierarchy Process (AHP) is based on the experience gained by its developer, T.L. Saaty [15], while directing research projects in the US Arms Control and Disarmament Agency. The simplicity and power of the AHP has led to its widespread use across multiple domains in every part of the world.

TOPSIS method is presented in Chen and Hwang [16], with reference to Hwang and Yoon [17]. The basic principle is that the chosen alternative should have the shortest distance from the ideal solution and the farthest distance from the negative-ideal solution.

Combining IGAPE method with AHP and TOPSIS in decision making provides adequate rationale for the decision arrived at.

Section 2 reviews various methods in GORE. Section 3 explains the basic steps of IGAPE. Section 4 illustrates the IGAPE method on an online shopping system. Section 5 discusses the results obtained.

## 2. RELATED WORK

Goal identification and refinement in the specification of Software-based information systems by Anton [18] proposed GBRAM method (Goal based requirements analysis and measurement).

Evangelia Kavakli's [19] work on modeling and guidance describes a Goal Driven Change (GDC) approach. He proposes a systematic way of reasoning about the RE process in terms of Goal modeling used within a process guidance framework.

Huzam Al-Subaie [20] work on evaluating the effectiveness of a GORE method makes an attempt to systematically evaluate the KAOS [7] (Knowledge acquisition in Automated Systems) method and the Objectiver tool using the main requirements engineering objectives.

NFR [6] proposes a comprehensive framework for representing and using non-functional requirements during the development process. The i* modeling framework is the basis for Tropos, a requirements-driven agent-oriented development methodology [21]. The framework has two main components: the Strategic Dependency (SD) model and the Strategic Rationale (SR) model.

In Attributed Goal-Oriented Requirements Analysis Method (AGORA) [22], an analyst attaches contribution values and preference values to edges and nodes of a goal graph respectively during the process for refining and decomposing the goals.

Visual Variability Analysis (VVA) for Goal models [23] proposes a visual technique for understanding requirements variability by providing an interactive visual analysis of variants.

A technique for specifying partial degrees of goal satisfaction and for quantifying the impact of alternative system designs on the degree of goal satisfaction is proposed by Emmanuel L. and Axel van Lamsweerde [24]. Axel van Lamsweerde [25] proposes a quantitative but lightweight technique for evaluating alternative options.

Paolo Giorgini et al [26] propose a formal framework for reasoning with goal models. Goal Argumentation Method (GAM) [27] guides argumentation and justification of modeling choices during the construction of goal model.

Comparative study of Goal-Oriented Requirements Engineering is undertaken in Shahzad Anwer and Naveed, Ikram,[28], Naeem, Ur Rehman et al [29]; Subhas M. et al [30] and Jennifer Horkoff and Eric Yu [31].

The key factor in any decision support system during requirements engineering is to identify relationship between functional and non-functional requirements. GORE methods do a good job in capturing these relationships.

GORE methods which make use of formal techniques like use of temporal logic and label propagation algorithms for decision support system have certain limitations:

1. Decision support techniques are not transparent to majority of stakeholders since majority of existing techniques are formal and qualitative in nature. Any project will have multiple stakeholders which include clients, customers, operators, requirement analysts, designers, testers and many more. The drawback of a formal approach is that stakeholders like clients, customers or operators will not be able to play an active role in decision support system since the process involved is too complex for them to understand. Many a times, decision taken is forced upon these stakeholders who end up playing very little role in decision support system.
2. Few techniques which are semi-formal lack sound reasoning on converting qualitative data into quantitative. The qualitative contribution links which exists between the hard and soft goals are interpreted in multiple ways by each method making it complex for all stakeholders to understand the principle behind it.
3. The knowledge base for decision support system in present GORE techniques is not sufficient enough for arriving at a decision based on sound rationale.
4. Not much work is carried out on involving multiple stakeholders whose perspectives on a given goal graph in terms of the contribution links between hard and soft goal might differ.

The limitations identified above serve as motivation to take up our research work. Decision support system in existing Goal Oriented requirements engineering are largely formal in nature and do not provide adequate scope for all the stakeholders to play an active role. We propose a semi-formal GORE method IGAPE to overcome this limitation.

The major contributions of our work include:
1. Identification of strategies for enhancing the knowledge base for decision support system.
2. Development of a framework for integrating IGAPE GORE method with AHP and TOPSIS.

# 3. INTEGRATION OF GOALS AFTER PRIORITIZATION & EVALUATION (IGAPE) METHOD

IGAPE method is shown in Fig 1 and the steps are described in Table 1.

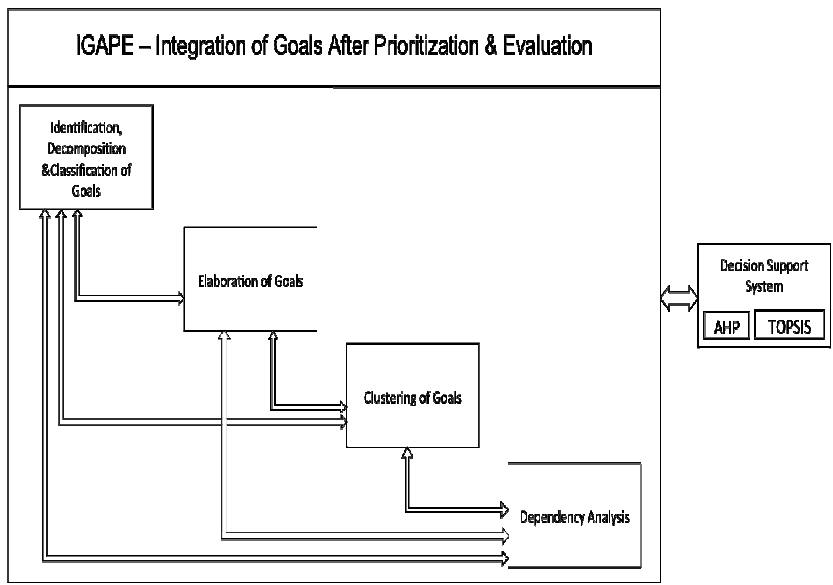

Figure 1. Integration of Goals after Prioritization & Evaluation (IGAPE) method.

Table 1 Steps involved in IGAPE

| Step | Description | Approach | Artefacts Produced |
|---|---|---|---|
| 1. | Identification, Decomposition and Classification of Goals | a) Identification: This involves identifying the goals for the application to be developed. These goals are identified by interacting with the various stakeholders and employing guidelines proposed by earlier GORE methods. | Goals, Functional Requirement (Hard Goal), Non-functional Requirement, Soft Goals |

| | | b) Decomposition: AND and OR decomposition of goals are done as per the needs of the application. | |
| | | c) Classification: A goal becomes either a functional requirement (hard goal) or a quality requirement only if it is a leaf level goal which cannot be decomposed further.<br>A quality requirement are of two types:<br>  1) Quantitative metric or Non-functional requirement which has a clear criteria for satisfaction.<br>  2) Qualitative metric or soft goal for which there are no a priori, clear criteria for satisfaction, but are judged by actors as being sufficiently met. Soft Goals are goals which do not have clear cut criteria to decide whether the goal is satisfied or not. | |
| 2. | Elaboration of Goals | Elaboration of identified goals using goal template. This template is of two types depending on the goal type:<br>  1) Leaf level goals<br>  2) Non-leaf level goals | Description of leaf level and non-leaf level goals as per the template. |
| 3. | Clustering of goals | Clustering mechanism is employed to reduce the complexity of the large number of goals. This will assist decision support system carried out in step 4. Guidelines for clustering are as follows:<br>a) All functional goals at the first level are candidate goals for becoming a root goal after clustering.<br>b) All functional goals at the first level of a clustered goal are candidate goals for becoming a root goal after clustering.<br>c) Step b) is applied on the need basis depending on the type of application being developed. | Clusters |
| 4. | Dependency Analysis | This involves identification of dependencies between hard goals and quality requirements in a Cluster. A hard goal can have contribution links to either of the two types of quality requirement. They can be:<br>  a) Contribution link between a hard goal and a non-functional requirement defined by a specific numeral. The non-functional requirement can be either a benefit attribute or negative | a) Goals with contribution links.<br>b) Updated templates of goals. |

|  |  | attribute. The objective is to maximize the benefit attribute and minimize the negative attribute. | |
|  |  | b) Contribution link (+, ++, o, -, --) between a hard goal and a soft goal (qualitative non-functional requirement). The meaning of these contribution links is as follows: | |

| Symbol | Meaning |
|---|---|
| ++ | A hard goal requirement fully supports the soft goal. |
| + | A hard goal requirement partially supports the soft goal. |
| o | Indicates neutral or no dependency between hard and soft goal. |
| - | A hard goal requirement minimally supports the soft goal. |
| -- | A hard goal requirement is not supported by the soft goal. |

| 5. | Decision Support System | This step involves identifying the scenarios in a goal graph and using AHP and TOPSIS to provide Decision support system. The scenarios can be of two types:<br>• Choosing an alternative among two or more functional goals to realize the root goal.<br>• Prioritizing and choosing the alternatives among the functional goals. | Ranking of Alternatives |

## 3.1 Decision Support System in IGAPE

This step involves identifying the scenarios in a goal graph and using AHP and TOPSIS to provide Decision support system. The scenarios can be of two types:

- Choosing an alternative among two or more functional goals to realize the root goal. The steps involved are shown in Table 2.
- Prioritizing and choosing the alternatives among the functional goals. The steps involved are shown in Table 3.

Table 2 Choosing an alternative among two or more functional goals.

| Step | Description | Method |
|---|---|---|
| 1 | Calculate priority of quality requirements | a) Use AHP to calculate priority of quality requirements. If it is a multi-level hierarchy, we need to proceed from root to leaf level goals in |

| | | calculating priorities of quality requirements using AHP. Local weights are calculated among same level goals using AHP.<br>b) Global weights are the product of all local weights proceeding from leaf to root. |
|---|---|---|
| 2 | Evaluation of each alternative using TOPSIS | a) Convert contribution link to numeric values in case of a soft goal.<br>b) Use the metric directly if it is a non-functional requirement.<br>c) Apply these numeric values for TOPSIS method while comparing each alternative for the identified quality requirements. |
| 3 | Ranking of alternatives | The alternatives are ranked in terms of increasing order of their values. |

Table 3. Prioritizing and choosing the alternatives among the functional goals.

| Step | Description | Method |
|---|---|---|
| 1 | Calculate priority of quality requirements | a) Use AHP to calculate priority of quality requirements. If it is a multi-level hierarchy, we need to proceed from root to leaf level goals in calculating priorities of quality requirements using AHP. Local weights are calculated among same level goals using AHP.<br>b) Global weights are the product of all local weights proceeding from leaf to root. |
| 2 | Evaluation of each alternative using TOPSIS | a) Convert contribution link (+, ++, o, -, --) to numeric values in case of a soft goal. We will be making use of QFD approaches (Akao Yoji [32]; Herzwurm G et al [33]; Andreas H et al [34]; De Felice and Petrillo A [35] for this purpose. A linear scale {2,1,0,-1,-2} is used for this purpose (Galster M et al [36])<br>b) Use the metric directly if it is a non-functional requirement.<br>c) Apply these numeric values for TOPSIS method while comparing each alternative for the identified quality requirements. |
| 3 | Ranking of alternatives | The alternatives are ranked in terms of increasing order of their values. |
| 4 | Calculate priority of functional goals | Use AHP to calculate priority of functional goals. |

| 5 | Choosing the alternatives | Depending on the output obtained in step 3 and 4, choose the alternatives among the functional requirements. |
|---|---|---|

### 3.1.1 AHP and TOPSIS

One of the important methods for prioritization is Analytic Hierarchy Process. We make use of AHP to prioritize hard goals and quality requirements (Non-functional goal and soft goal). An evaluation of six methods for prioritizing software requirements suggests AHP as the promising method for prioritizing requirements (Joachim Karlsson et al [37]). The study found that even though AHP process is demanding but is worth the effort because of its ability to provide reliable results, promote knowledge transfer and create consensus among project members.

The output of IGAPE comprises quality requirements which are quantitative (non-functional requirement) and qualitative (soft goal). We needed to choose a method which can take both quantitative and qualitative data as input. TOPSIS method can combine these two kinds of data effectively. One of the recent works on value based requirement prioritization by Kukreja N et al [38] compared 17 frameworks used for requirement prioritization. 17 criteria which include scalability, sensitivity analysis, scientific credibility and many more were chosen for comparison of various requirements prioritization techniques. The study concluded that TOPSIS satisfied most of the criteria to the fullest and stood as the best with respect to the 17 criteria and their relative weights.

## 3.2 Goal Description Template

The identified goals are elaborated as per the template depending on whether a particular goal is a leaf goal or non-leaf goal. The template for non-leaf goal is as follows:

### 3.2.1 Non-leaf Goal Template

- Goal ID: Unique Identifier of the goal
- Cluster ID: The unique identifier of the cluster to which the goal belongs to.
- Goal Name: Unique name for the goal
- Authors: Name of the authors who have documented the goal
- Goal Description: Description of the goal
- Source: Name of the source (i.e. stakeholder, document or system) from which the goal originates and the rationale.
- Super Goal ID: Reference to the super goal including the type of decomposition (AND/OR)
- Sub-goal/s ID: Reference to the sub-goals including the type of decomposition (AND/OR)
- Version: Current version number of the documentation of the goal
- Change History: List of the changes applied to the documentation of the goal including (for each change) the date of change, the version number, the author, and, if necessary, the reason for and the subject of change.
- References: Any other relevant information about the goal or adequate references

### 3.2.2 Leaf Goal Template

- Goal ID: Unique Identifier of the goal
- Goal Name: Unique name for the goal
- Goal Type: Goal type can be either a
  - o Functional Requirement (Hard Goal) or
  - o Non-Functional Benefit Requirement (NFRB)
  - o Non-Functional Negative Requirement (NFRN)
  - o Soft Goal
- Authors: Name of the authors who have documented the goal
- Goal Description: Description of the goal
- Source: Name of the source (i.e. stakeholder, document or system) from which the goal originates and the rationale.
- Stakeholders: Stakeholders who benefit from the satisfaction of the goal.
- Assignment: Name of the stakeholder who is responsible for the goal.
- Super Goal ID: Reference to the super goal including the type of decomposition (AND/OR)
- Sub-goal/s ID: Reference to the sub-goals including the type of decomposition (AND/OR)
- Contribution links: The contribution links can be of two types:
  - o Functional requirement to a Non-functional requirement (NFR)
  - o Functional requirement to a Soft goal (SG)

| Contribution Link Type (NFR / SG) | From (Goal ID) | To (Goal ID) | Metric (if it is NFR) | Value (if it is NFR) | Qualitative (if it is SG) |
|---|---|---|---|---|---|
|  |  |  |  |  |  |
|  |  |  |  |  |  |

- Dependencies: Requires/Conflict
  - o Requires: A "requires" dependencies defines that, to satisfy $G_1$, the goal $G_2$ must be satisfied.
  - o Conflict: A "conflict" dependency represents that the satisfaction of one goal hinders the satisfaction of the other goal.

| Dependency Type (Requires / Conflict) | From(Goal ID) | To (Goal ID) | Description |
|---|---|---|---|
|  |  |  |  |
|  |  |  |  |

- Acceptance Criteria (for FR): The acceptance criteria to ascertain whether the objective of the goal has been realized or not.
- Obstacle Analysis – Identify scenarios which prevent realization of the goal and come up with resolutions for each of the scenario(for FR)
- Attribute: Attributes can be of the following types:

- o Stability: Stability is a measure for the likelihood that the associated requirement will change in the course of the project. Range of values: { volatile; presumably stable; stable }
- o Negotiation Status: Describes the current level of agreement about the artefact. Range of values: { unknown, conflicting, in-agreement, agreed }
- o Priority: Describes the importance of the requirement for achieving the overall goals defined for the system> Range of values: {high, medium, low}
- Version: Current version number of the documentation of the goal
- Change History: List of the changes applied to the documentation of the goal including (for each change) the date of change, the version number, the author, and, if necessary, the reason for and the subject of change.
- References: Any other relevant information about the goal or adequate references.
- Use Case ID (for FR): Refers to the use case ID for the functional requirements.

# 4. CASE STUDY ILLUSTRATION

We illustrate the steps proposed in our method by considering an Online Shopping System application.

## 4.1 Identification, Decomposition and Classification of goals

The goals identified at the first level include (we have omitted few of the goals due to space constraints):

- Good User Interface
- Search Facility
- Attractive price on products
- Payment system
- Delivery system
- Support System

We will illustrate the IGAPE method for Payment System and Support System goals.

## 4.2 Payment System

- Payment System
    - o Payment Gateway (Credit Card, Debit Card, Net Banking)
        - ▪ Option A
        - ▪ Option B
        - ▪ Option C
        - ▪ Option D
    - o Mobile Payment
    - o Cash Cards
    - o Online Bank Transfer
- Cash on Delivery

We will identify Payment Gateway as a Cluster and continue exploration of goals. There are 4 options to choose from. We will identify quality requirements which will play an important role in deciding the best alternative.

**4.2.1 Quality Requirements for choosing the Payment Gateway**

The quality requirements necessary for choosing the payment gateway is shown in Table 4. It comprises both non-functional requirements (Goals numbered 1.1-1.3, 2 and 3.1-3.4) and soft goal (goals numbered 4.1-4.3). The table also depicts the relationship between goals and four alternatives.

Table 4. Prioritizing and choosing the alternatives among the functional goals.

| Level 1 | Level 2 | Type of Quality Requirement | Option D | Option C | Option B | Option A |
|---|---|---|---|---|---|---|
| 1. Cost | 1.1 Set up Fee | NFRN | 7500 | 50000 | 30000 | 50000 |
| | 1.2 Transaction Discount Rate | NFRN | 5.5 | 5 | 4 | 4 |
| | 1.3 Annual Maintenance charges | NFRN | 1200 | 12000 | 0 | 15000 |
| 2. Number of Credit Card Support | | NFRB | 6 | 2 | 2 | 3 |
| 3. Technical | 3.1 Unscheduled Down Time (in hours per month) | NFRN | 2 | 1 | 1 | 2 |
| | 3.2 Transaction Success Rate (% per 1000 transactions) | NFRB | 75.3 | 73.52 | 78.37 | 58.84 |
| | 3.3 Average Time taken for refunds in Days | NFRN | 5.5 | 4.03 | 3.9 | 3.45 |
| | 3.4 Set up Time | NFRN | 15 | 20 | 12 | 15 |
| 4. Qualitative | 4.1 Customer Service | 4.1.1 Support System Availability (SG) | ++ | + | + | + |
| | | 4.1.2 Dispute Resolution (SG) | + | + | + | - |
| | 4.2 Customer Trust (SG) | | + | + | + | - |

| | 4.3 Ease of Integration (SG) | | ++ | + | + | + |
|---|---|---|---|---|---|---|

## 4.2.2 Decision Support System

The objective is to choose the best payment gateway for the quality factors identified. The steps specified in Table 2 are applied.

Step 1: Calculation of priorities of Quality requirements is done using AHP. The local and global values are shown in Table 5. Global weights are the product of local weights.

Table 5. Calculation of Global weights using AHP

| Level 1 | Local Weights | Level 2 | Local Weights | Level 3 | Local Weights | Global Weights |
|---|---|---|---|---|---|---|
| Cost | .235 | Set up Fee | .221 | - | - | 0.0519 |
| | | Transaction Discount Rate | .451 | - | - | 0.1059 |
| | | Annual Maintenance charges | .328 | - | - | 0.0770 |
| Number of Credit Card Support | .125 | | | - | - | .125 |
| Technical | .442 | Unscheduled Down Time (in hours per month) | .112 | - | - | 0.0495 |
| | | Transaction Success Rate (% per 1000 transactions) | .523 | - | - | 0.2311 |
| | | Average Time taken for refunds in Days | .256 | - | - | 0.1131 |
| | | Set up Time | .109 | - | - | 0.0481 |
| Qualitative | .198 | Customer Service | .442 | Support System Availability (SG) | .656 | 0.0574 |
| | | | | Dispute Resolution (SG) | .344 | 0.0301 |
| | | Customer Trust (SG) | .335 | - | - | 0.0660 |
| | | Ease of Integration | .223 | - | - | 0.0441 |

| | | (SG) | | | | | | | | | |
|---|---|---|---|---|---|---|---|---|---|---|---|

Step 2: Evaluation of each alternative using TOPSIS. The input to TOPSIS is shown in table 6. The global weights computed for all quality requirements in table 5 are also part of the input to the TOPSIS technique.

Table 6. Input to TOPSIS

| QR ID/ Alternative | 1.1 | 1.2 | 1.3 | 2 | 3.1 | 3.2 | 3.3 | 3.4 | 4.1.1 | 4.1.2 | 4.2 | 4.3 |
|---|---|---|---|---|---|---|---|---|---|---|---|---|
| Option D | 7500 | 5.5 | 1200 | 6 | 2 | 75.3 | 5.5 | 15 | 1 | 1 | 1 | 2 |
| Option C | 50000 | 5 | 12000 | 2 | 1 | 73.52 | 4.03 | 20 | 1 | 1 | 1 | 1 |
| Option B | 30000 | 4 | 0 | 2 | 1 | 78.37 | 3.9 | 12 | 1 | 1 | 1 | 1 |
| Option A | 50000 | 4 | 15000 | 3 | 2 | 58.84 | 3.45 | 15 | 1 | -1 | -1 | 1 |

The output of TOPSIS is shown in Graph 1. We can infer the ranking of alternatives as Option D, Option B, Option C and Option A from the graph. Option D is the most preferred payment gateway which fulfills most of the quality requirements.

Graph 1 Output of TOPSIS

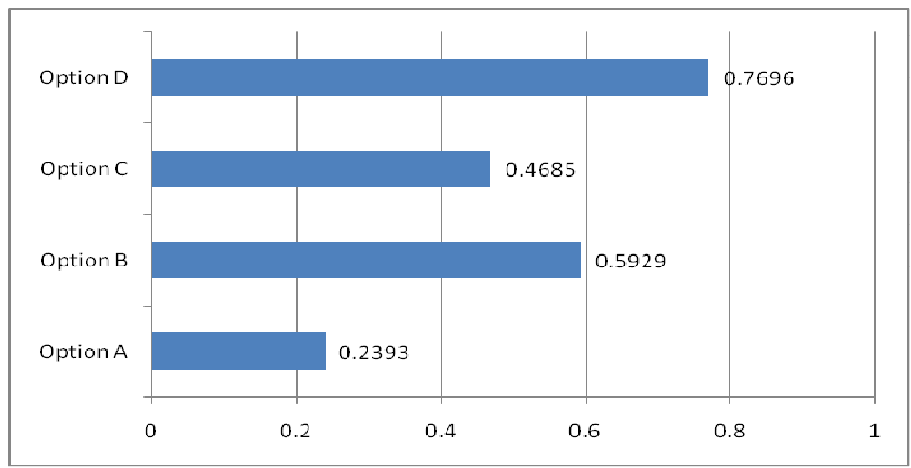

## 4.3 Support System

Providing good support system is one of the important goals of an e-commerce website.
- Support System
  - Product Information
    - Online Chat
    - Telephone (Toll-free)
    - Email
  - Purchase Support
    - Online Chat
    - Telephone (Toll-free)
    - Email
  - General Feedback
    - Online Chat

- Telephone (Toll-free)
- Email

The objective here is to prioritize and choose all or few of the alternatives among the goals. The steps mentioned in Table 3 are applied for this scenario. First we prioritize among the three sub-goals of support system: Product Information, Purchase Support and General Feedback. This is done using AHP. The output of this step is shown in Table 7.

Table 7. Prioritization using AHP among Goals

| Goal | Global Weight |
|---|---|
| Product Information | 0.255 |
| Purchase Support | 0.520 |
| General Feedback | 0.225 |

### 4.3.1 Quality Requirements for choosing a specific support system

All three goals i.e. Product Information, Purchase Support and General Feedback can be realized by any of the following options:

- Online Chat
- Telephone (Toll-free)
- Email

We evaluate which among the following three options are best suited. The quality requirements for determining it are as follows:

a) Amount of training needed in days: NFRN
b) Effort needed for Feedback monitoring of one support staff in days: NFRN
c) Cost needed for setting up required Infrastructure : NFRN
d) Time taken for resolving customer query satisfactorily in hours: NFRN
e) Customer Convenience: SG

We apply AHP to prioritize among the above mentioned quality requirements. The output of this is shown in table 8. Table 9 shows the input to TOPSIS. Graph 2 shows the output of TOPSIS.

Table 8. Prioritization using AHP among quality requirements

| Goal ID | Goal | Global Weight |
|---|---|---|
| a | Amount of training needed | 0.178 |
| b | Effort needed for Feedback monitoring | 0.120 |
| c | Cost | 0.125 |
| d | Time taken for resolving query | 0.350 |
| e | Customer Convenience | 0.227 |

Table 9. Input to TOPSIS

| QR ID/ Alternative | (a) | (b) | (c) | (d) | (e) |
|---|---|---|---|---|---|
| Online Chat | 3 | 2 | 2000 | 4 | 1 |
| Telephone (Toll-free) | 4 | 3 | 3000 | 3 | 2 |
| Email | 2 | 3 | 2000 | 5 | 1 |

Graph 2 Output of TOPSIS

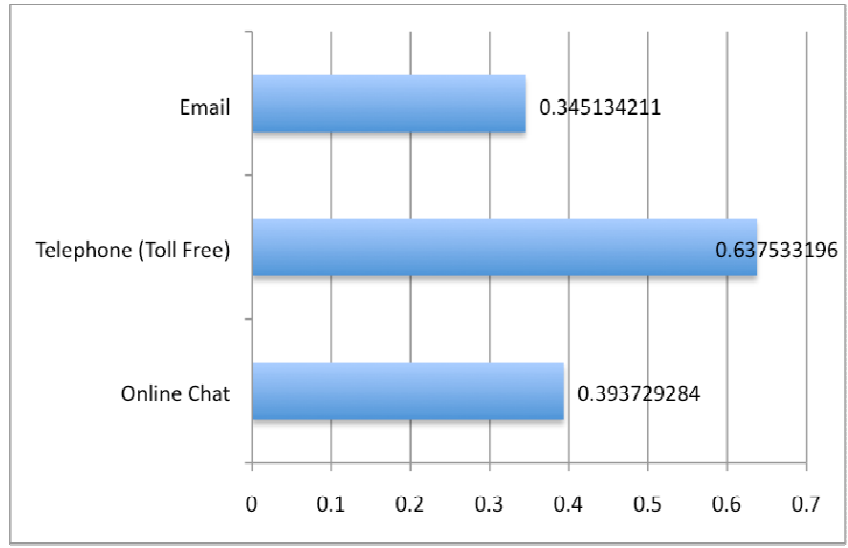

We can infer the ranking of alternatives as Telephone (Toll free), Online Chat, Email from graph 2. Depending on the prioritized values as shown in Table 7, we can choose either all or few or any one among the alternatives as shown in Table 10.

Table 10. Prioritization using AHP among Goals

| Goal | Global Weight | Alternatives Chosen |
|---|---|---|
| Product Information | 0.255 | Telephone (Toll free) |
| Purchase Support | 0.520 | Telephone (Toll free), Online Chat, Email |
| General Feedback | 0.225 | Email |

# 5. RESULTS AND DISCUSSION

The approach to validate the proposed method of combining AHP, TOPSIS and GORE is done by considering the opinion of the experts. This approach can be found in the work of M. Galster et al [36]. The guidelines followed are as follows:

- Problem Description of Payment Gateway and Support system were given to experts.
- Experts had to recommend one choice among the alternatives.
- Experts had to reject those choices which were not appropriate.
- Experts had to rank the alternatives.

- The experts will use their own approach in deciding the alternative and are not aware about our proposed method.
- Comments regarding their decision may be provided.
- All experts are from Industry had 5+ years experience and at least have an undergraduate degree in Computer Science.

The comparison of the decision taken by experts with the results obtained by our approach for payment gateway system is shown in Table 11. Alternatives are as follows:

- A1 – Option D
- A2 - Option B
- A3 - Option C
- A4 - Option A

Table 11. Comparison of our method with experts

| Subject | Chosen alternative | Rejected alternative | Ranking of alternatives |
|---|---|---|---|
| Our method | A1 | A4 | A1, A3, A2, A4 |
| Expert 1 | A1 | A4 | A1, A2, A3, A4 |
| Expert 2 | A2 | A4 | A2, A1, A3, A4 |
| Expert 3 | A1 | A4 | A1, A3, A2, A4 |
| Expert 4 | A3 | A4 | A3, A1, A2, A4 |
| Expert 5 | A1 | A4 | A1, A3, A2, A4 |
| Expert 6 | A1 | A4 | A1, A3, A2, A4 |

Results shown in Table 10 clearly convey that A1 – Option D is chosen as the best alternative by four experts as well as our approach. One expert has chosen A2 and the other A3. Three experts have the same ranking of alternatives as our method. The experts have arrived at a decision based on their domain knowledge, importance of quality requirements and in consultation with other colleagues. We can clearly infer that our method results are in agreement with the experts. The advantage of our method is that it provides a precise metric and rationale for the decision arrived at using AHP and TOPSIS unlike experts who rely more on experience and intuition. Another added benefit is that all the stakeholders can play an active role in decision support.

## 5.1 Kendall's Coefficient of Concordance

Kendall's Coefficient of Concordance [39] provides a mechanism to study the degree of association among three or more sets of rankings. This descriptive measure of the agreement has special applications in providing a standard method of ordering objects according to consensus when we do not have an objective order of objects.
Find coefficient of concordance (W) which is an index of divergence of the actual agreement shown in the data from the perfect agreement. This is calculated as follows

a) All the alternatives (N=4), should be ranked by all k experts including our method (k=7) and this information may be put in the form of a k by N matrix. This is shown in table 12.

Table 12 Comparison of our method with experts

| Expert / Alternatives | A1 | A2 | A3 | A4 |
|---|---|---|---|---|
| **Our method** | 1 | 3 | 2 | 4 |
| **Expert 1** | 1 | 2 | 3 | 4 |
| **Expert 2** | 2 | 1 | 3 | 4 |
| **Expert 3** | 1 | 3 | 2 | 4 |
| **Expert 4** | 2 | 3 | 1 | 4 |
| **Expert 5** | 1 | 3 | 2 | 4 |
| **Expert 6** | 1 | 3 | 2 | 4 |

b) For each alternative determine the sum of ranks $(R_j)$ assigned by all stakeholders

c) Determine $R_j^1 = \sum R_j / N$ and then obtain the value of $s$ as under

$$s = \sum ( R_j - R_j^1 )^2$$

d) Work out the value of W using the following formula

$$W = s / (( k^2 / 12) ( N^3 - N))$$

When perfect agreement exists between judges, W is equal to 1. When maximum disagreement exists, W is equal to 0. If the value of W is >=0.70 it indicates good level of agreement among stakeholders, then the best estimate of the true ranking is provided by the order of the sum of the ranks. The best alternative is related to the lowest value observed amongst $R_j$. Table 13 shows computation of s value.

Table 13. Computation of *s* value

| Expert / Alternatives | A1 | A2 | A3 | A4 | |
|---|---|---|---|---|---|
| **Our method** | 1 | 3 | 2 | 4 | |
| **Expert 1** | 1 | 2 | 3 | 4 | |
| **Expert 2** | 2 | 1 | 3 | 4 | |
| **Expert 3** | 1 | 3 | 2 | 4 | |
| **Expert 4** | 2 | 3 | 1 | 4 | |
| **Expert 5** | 1 | 3 | 2 | 4 | |
| **Expert 6** | 1 | 3 | 2 | 4 | |
| **Sum of Ranks ($R_j$)** | 9 | 18 | 15 | 28 | $\sum R_j = 70$ |
| $( R_j - R_j^1 )^2$ | 72.25 | .25 | 6.25 | 110.25 | $s = 189$ |

Finally W is calculated using equation specified in step (d) with values of k=7, s= 189 and N=4. We get W= 0.771 which indicates good agreement on ranking among the experts and our method. The best alternative is related to the lowest value observed amongst $R_j$ which is Alternative A1 – Option D.

In the existing GORE literature, there exists technique which makes use of formal techniques [7]. They make use of temporal logic and label propagation algorithms. Our approach differs in adopting a quantitative way of evaluating the alternative using AHP. We believe that a quantitative approach will enable all stakeholders to play an active role in decision making.

One of the existing methods which make use of a quantitative approach for decision support system in GORE is Attributed Goal Oriented Requirements Analysis (AGORA) [22]. The future works suggested by the authors mention use of AHP to assign the values subjectively. We have used AHP and TOPSIS. Combining our GORE method with AHP and TOPSIS in decision making provides adequate rationale for the decision arrived at.

## 5.3 Our Contribution

The major contributions of our work are as follows:

- **Enhancing the Knowledge base in IGAPE method:** Following are the unique features of our IGAPE method which is not present in any of the existing GORE methods:
  - o Identification of quality requirements which are of three types: Most of the methods use soft goals or non-functional requirement but our approach makes use of non-functional requirement which can be either of benefit type or negative type and soft goals. The presence of benefit or negative type makes the task of either optimizing (maximizing or minimizing) a specific NFR which will help us in choosing the best alternative or prioritizing among goals easier.
  - o Clustering Mechanism: Most of the existing GORE methods find it difficult to manage the goals as it grows in number. IGAPE method makes use of the concept of Clusters to manage them in a better way. A cluster will have its own quality requirements along with the functional goals. This will make the task of decision making easier.
  - o Use of quantitative and qualitative attribute: IGAPE method combines the use of quantitative attribute (NFRB or NFRN) and qualitative attribute (SG) in Decision support system.
  - o Analyzing scenarios in a Goal graph: By making use of the Clustering concept we have identified two scenarios which involves:

    - ▪ Evaluation of alternatives to realize the root goal.
    - ▪ Prioritization of goals and choosing one or few or all the alternatives which facilitate the realization of root goal in a cluster.

- **Active stakeholder involvement in Decision making:** IGAPE method facilitates active stakeholder involvement which may not be possible with other formal GORE decision support systems. These formal techniques need an expert analysis which makes it difficult for all stakeholders to play an engaging role.

- **IGAPE is semi-formal technique:** The major limitation of decision support in GORE method is the lack of semi-formal or quantitative decision support technique. To overcome this we have proposed a GORE method which makes use of AHP and TOPSIS.

- **Use of AHP and TOPSIS methods in GORE:** AHP and TOPSIS have not been used in any GORE method. The IGAPE method combines the benefits of GORE with AHP and TOPSIS methods.

We have identified limitations of KAOS and AGORA method in our literature survey. Decision support techniques in requirements engineering are not aimed at obtaining optimized solution. The effort is to provide a knowledge base to enable decision support rather than decision making.

Hence it is not possible to precisely validate the correctness of the decision made. However the proposed method provides proper reasoning for the decision made. The use of AHP and TOPSIS is specified and as explained earlier the correctness is dependent on the quality of the goal model.

In addition to AHP, we make use of TOPSIS which adds more rigors to decision support. TOPSIS can effectively combine quantitative and qualitative data.

### 5.4 Tool support

We have developed a software tool for IGAPE method. The tool has been developed using C# .NET. The major features of the tool include:

- Support for entering details about hard goals, soft goals and NFR specific to an application.
- Provision for documenting all decomposed goals and dependencies among goals.
- Calculation of local and global weights using AHP.
- Final ranking of alternatives using TOPSIS method.

### 5.5 Scope for future work

The work on integrating AHP and TOPSIS into IGAPE method is presented. The future work will be carried out in the following areas:

(i) Different stakeholders will have their own perspective while prioritizing quality requirements or hard goals. The method needs to factor this in decision support system.

(ii) Identifying stakeholders relevant to a specific cluster and according them priority needs to addressed.

(iii) Exploring possibility of including game theoretic approaches to decision support system in GORE.

## 6. CONCLUSION

We have proposed IGAPE method which is based on the principles of goal-oriented requirements engineering. The method enhances the knowledge base by identifying relationship between functional and quality requirements which is vital for a good decision support system in requirements engineering. The semi-formal method ensures active stakeholder participation. We have elaborated the various steps of IGAPE method. The method makes use of Analytic Hierarchy Process (AHP) and Technique for Order of Preference by Similarity to Ideal Solution (TOPSIS) which are used in the industry. Integration of IGAPE with AHP and TOPSIS provides a rationale for various decisions which are arrived at during the requirements engineering phase. A software tool to support the IGAPE method is also developed.

**Authors**

**Vinay S** did his post graduation from SJCE, Mysore and is presently pursuing octoral studies at Manipal University. He is working as Associate Professor and Head in the Department of Computer Science and Engineering at PESITM, Shimoga. His main research interests are requirements engineering, decision support system, cloud computing and big data. He is a member of IEI, India and IAENG.

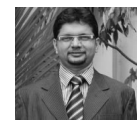

**Shridhar Aithal** received PhD in Energy Sciences, Masters in Engineering Physics from Canada. He holds Diploma in Computer Systems Technology and PMI certified PMP (Project Management Professional). He has over two decades of experience working in High Tech Industry in a leadership role (Project management in Technical Product development projects involving design, development, technical literature generation, testing, manufacturing, characterization, FCC /CSA/EU certification) experience. He has also worked as an academician in reputed institutions for over a decade.

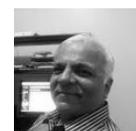

**Sudhakara G** received his Ph.D in Graph theory and is presently working as Professor in Department of Mathematics, MIT, Manipal. His research interests are graph theory, algebra and combinatorics. He is a life member of Ramanujan Mathematical Society.

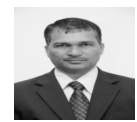

INTENTIONAL BLANK